\documentclass{ptapap}

\usepackage{amssymb}
\usepackage{url}

\author{Mariusz Tarnopolski}[NCU]
\affil[NCU]{Institute of Astronomy, Faculty of Physics, Astronomy and Informatics, Nicolaus Copernicus University, ul. Grudzi\k adzka 5, PL-87-100 Toru\'n, Poland}

\title{Quasi-periodic oscillations in gamma-ray bursts' prompt light curves}

\begin{document}

\maketitle

\begin{abstract}

I report on the discovery of 34 new quasi-periodic oscillations (QPOs) in the prompt light curves of long gamma-ray bursts (GRBs) from the Swift/BAT catalog: with one or more constant leading periods, as well as several chirping signals. This is the largest homogenously identified sample or GRB QPOs to date. The presence of QPOs suggests the existence of characteristic time scales that at least in some GRBs might be related to the dynamical properties of plasma trajectories in the accretion disks powering the relativistic jets. Several scenarios for their origin were examined. We identify non-planar orbits around Kerr black holes, the Lense-Thirring effect, and shock oscillations as plausible mechanisms for the QPO generation.

\end{abstract}

\section{Introduction}

The first quasi-periodic oscillation (QPO) in a prompt light curve (LC) of a gamma-ray burst (GRB) was observed in the 1979 March 5 event \citep{barat79,terrell80} -- an unambiguous $\sim 8$~sec modulation lasting about 20 periods. Some QPO candidates were reported in the following decades (see \citealt{tarnopolski21} for an overview), but GRB 090709A, observed by several space-borne gamma-ray observatories, gained attention when a rather prominent $\sim 8$~sec periodicity was noticed in its prompt LC. The significance varied, from below 3$\sigma$ to about 3.5$\sigma$  \citep{deluca10,cenko10,iwakiri10,ziaeepour11}, depending on the instrument and particular analysis method employed. A possible origin scenario might be magnetorotational instabilities in a hyperaccreting disk \citep{masada07}.

Herein, I report on the discovery of 34 new QPOs in the Swift/BAT catalog of long GRBs using a wavelet-based approach. This work is based on \citep{tarnopolski21}.

\section{Methodology}

The prompt LCs of GRBs in a 64~ms binning from the Swift/BAT catalog\footnote{\url{https://swift.gsfc.nasa.gov/results/batgrbcat/}} of $>$1300 GRBs were investigated. Selected were GRBs with duration $T_{100}>3.2$~sec, i.e., with at least 51 points in the LCs. Confirmed short GRBs with extended emission\footnote{\url{https://swift.gsfc.nasa.gov/results/batgrbcat/summary\_cflux/summary\_ GRBlist/GRBlist\_short\_GRB\_with\_EE.txt}} were excluded. The final sample yielded 1160 GRBs. 

QPOs were searched for in wavelet scalograms. A wavelet is a wave packet, i.e., an oscillation with a given frequency localized in time. The method implemented in the package \textsc{wavepal}\footnote{\url{https://github.com/guillaumelenoir/WAVEPAL}} \citep{lenoir18a,lenoir18b} was utilized. To test the significance of the detected features, they were tested against a continuous-time autoregressive moving average stochastic model \citep{kelly14}. A QPO was claimed if it lasted at least three cycles and exceeded the 3$\sigma$ significance throughout its duration.

\section{Results}

\begin{figure}
  \centering
    \includegraphics[width=\textwidth]{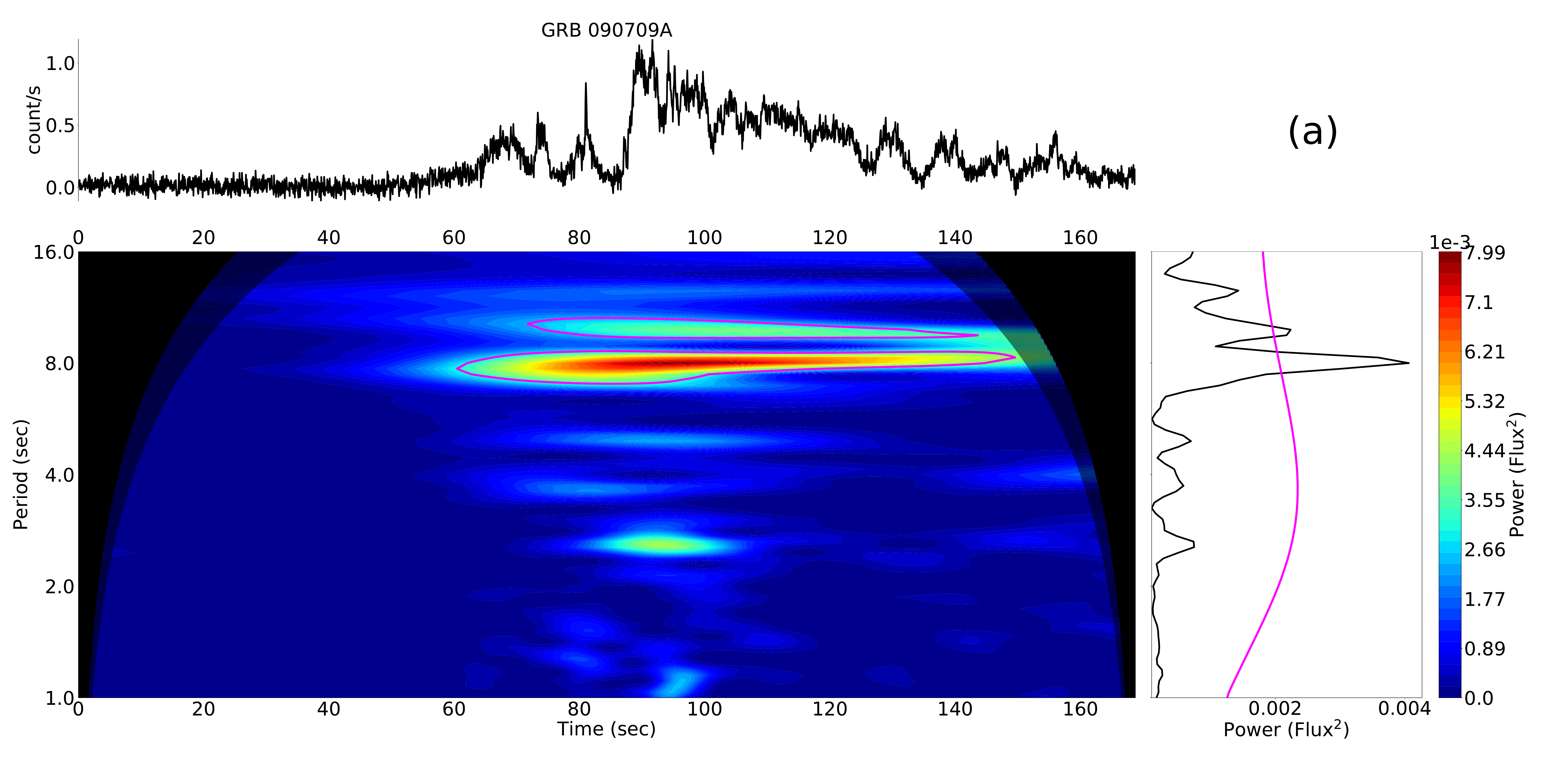}
    \caption{A wavelet scalogram of GRB 090709A confirmed the previously claimed QPO with a period of $\sim 8$~sec, but revealed also another, simultaneous modulation with a period of $\sim 9.8$~sec.
    Plot after \citep{tarnopolski21}. \textcopyright\, AAS. Reproduced with permission.}
    \label{fig1}
\end{figure}

Among the QPOs detected for 34 GRBs, 24 had at least one leading period: 13 GRBs exhibited one, 8 GRBs had two, and 3 had three coexisting QPOs (the latter two cases were denoted 'harmonics' in Table~\ref{table1}). Fig.~\ref{fig1} confirms the $\sim 8$~sec QPO in GRB 090709A; it reveals also another component with an $\sim 9.8$~sec period. These modulations were overlaid on a fast-rise-exponential-decay pulse.

In 10 GRBs a chirping signal was observed, i.e., the frequency was evolving in time. In cases when the frequency increases (period decreases) it is called an up-chirp. Similarly, when the frequency decreases (period increases) it is a down-chirp.

\begin{figure}
  \centering
    \includegraphics[width=\textwidth]{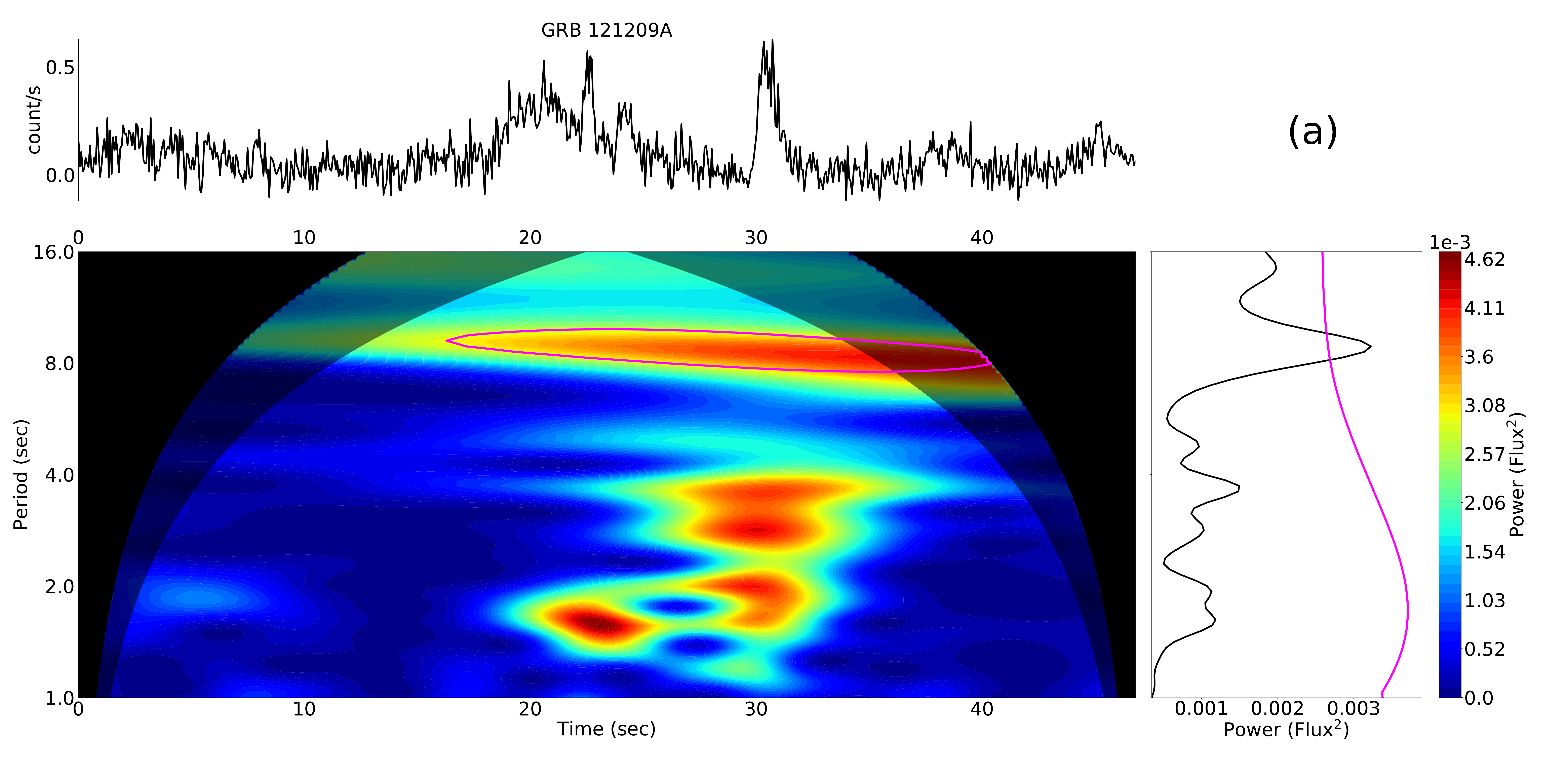}
    \includegraphics[width=\textwidth]{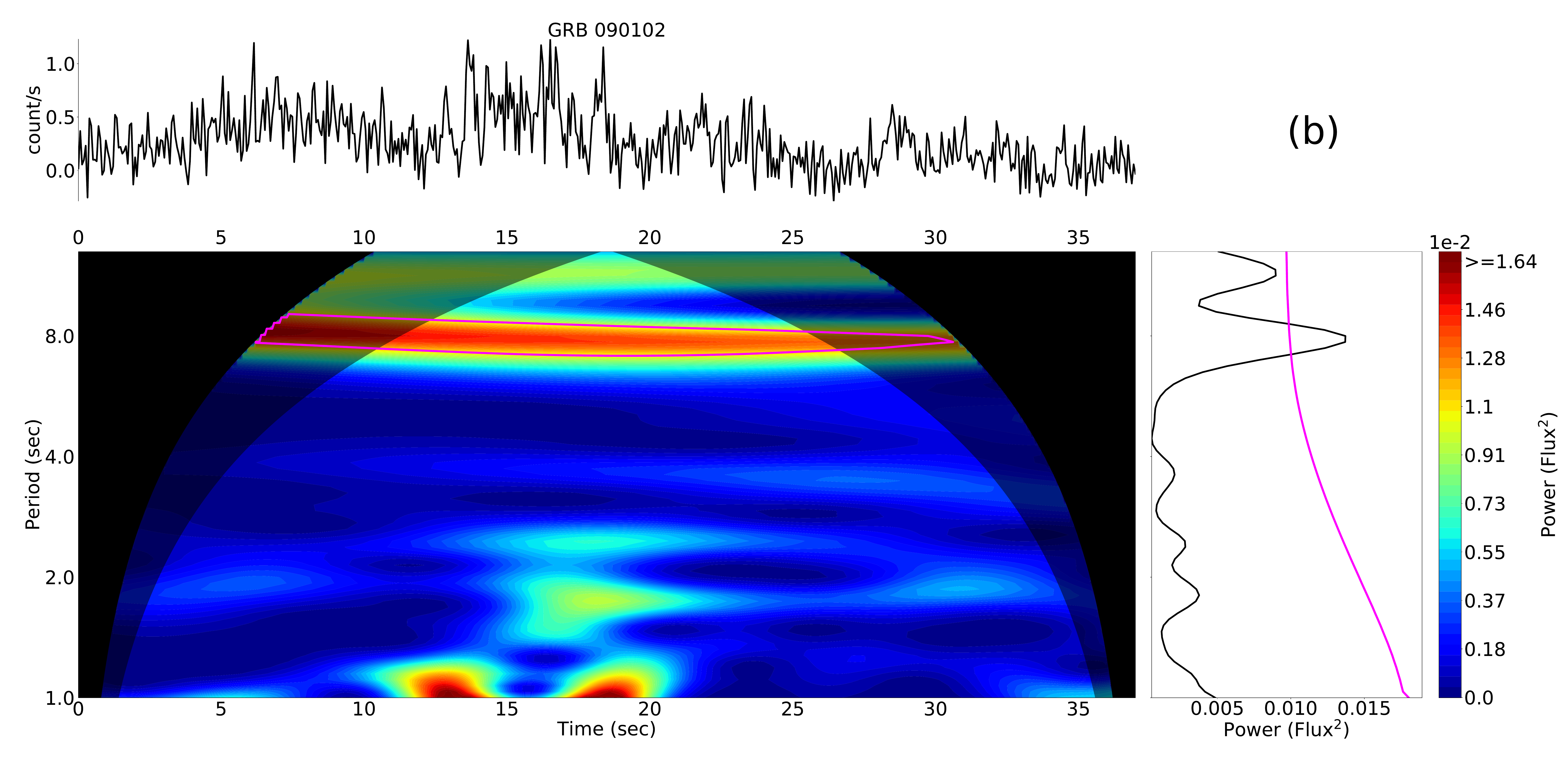}
    \includegraphics[width=\textwidth]{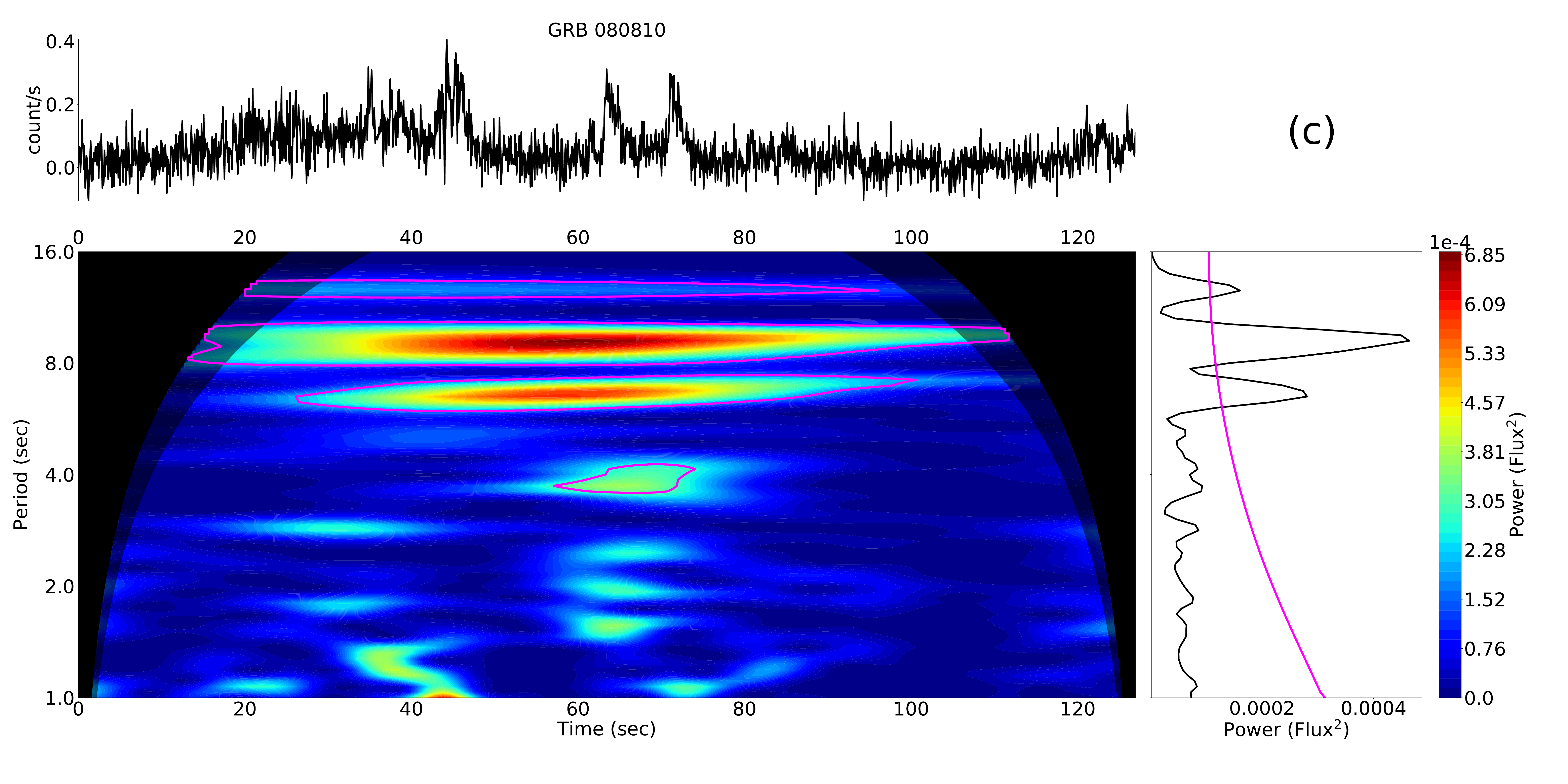}
    \caption{Eamples of (a) an up-chirp, (b) a constant leading period, and (c) peculiar harmonics in an apparent $2:3:4$ resonance. Cf. Table~\ref{table1}. Plot after \citep{tarnopolski21}. \textcopyright\, AAS. Reproduced with permission.}
    \label{fig2}
\end{figure}

Fig.~\ref{fig2} shows examples of novel detections of a slightly chirping signal (GRB 121209A), another constant leading period (GRB 090102), and a peculiar case of harmonics remaining in an apparent $2:3:4$ ratio (GRB 080810).

\begin{table}
\scriptsize
\begin{center}
\caption{Identified QPOs. Data after \citep{tarnopolski21}. \textcopyright\, AAS. Reproduced with permission. \textbf{Notes.} Approximately constant leading periods are given with corresponding uncertainties (indicated with the '$\pm$' sign). Period ranges of the chirping signals are indicated with arrows, '$\rightarrow$', showing the direction of period evolution. For the harmonics, the closest integer ratios are provided.\\$^*$These high-order ratios might as well be spurious, or be obscured due to uncertainties.}
\begin{tabular}{cccc}
\hline
Number & GRB Name & Period (sec) & Comment \\
\hline
6    & GRB200107B & $7.49\pm 1.16$; $11.40\pm 1.57$ & harmonics, $2:3$ \\
34   & GRB190821A & $8.20\rightarrow 5.28$ & up-chirp \\
75   & GRB190103B & $5.17\pm 0.76$ & constant \\
102  & GRB180823A & $19.12\pm 3.19$ & constant \\
122  & GRB180626A & $4.58\pm 0.33$; $5.70\pm 0.54$ & harmonics, $4:5$ \\
190  & GRB170823A & $2.96\rightarrow 11.58$ & down-chirp \\
212  & GRB170524B & $2.1\rightarrow 2.8$ & down-chirp \\
232  & GRB170205A & $6.86\pm 0.80$ & constant \\
250  & GRB161202A & $24.27\rightarrow 16.25$ & up-chirp \\
251  & GRB161129A & $3.83\rightarrow 6.95$ & up-chirp \\
252  & GRB161117B & $3.82\pm 0.52$ & constant \\
272  & GRB160824A & $3.05\pm 0.56$; $5.37\pm 0.81$; $9.43\pm 1.51$ & harmonics, $4:7:12^*$ \\
455  & GRB140730A & $12.32\pm 1.95$ & constant \\
462  & GRB140709B & $20.90\pm 2.00$; $41.54\pm 4.30$ & harmonics, $1:2$ \\
470  & GRB140619A & $8.87\pm 0.99$; $13.10\pm 1.85$; $32.34\pm 3.86$ & harmonics, $6:15:22^*$ \\
496  & GRB140323A & $5.49\pm 0.98$; $21.31\pm 2.91$ & harmonics, $1:4$ \\
551  & GRB130812A & $2.26\pm 0.40$ & constant \\
618  & GRB121209A & $9.89\rightarrow 7.57$ & up-chirp \\
622  & GRB121125A & $4.29\pm 0.73$; $8.48\pm 1.00$ & harmonics, $1:2$ \\
632  & GRB121014A & $16.70\pm 1.87$ & constant \\
701  & GRB120116A & $8.16\pm 0.96$ & constant \\
756  & GRB110422A & $5.46\rightarrow 3.89$ & up-chirp \\
777  & GRB110207A & $6.26\pm 0.74$ & constant \\
783  & GRB110107A & $5.48\rightarrow 3.46$ & up-chirp \\
805  & GRB100924A & $20.18\rightarrow 5.14$ & up-chirp \\
914  & GRB090709A & $8.02\pm 0.67$; $9.80\pm 0.91$ & harmonics, $4:5$ \\
945  & GRB090404 & $10.94\pm 0.86$ & constant \\
963  & GRB090102 & $7.64\pm 1.07$ & constant \\
1007 & GRB080810 & $6.70\pm 0.60$; $9.15\pm 0.85$; $12.67\pm 0.81$ & harmonics, $2:3:4$ \\
1098 & GRB070911 & $4.97\pm 0.75$; $16.50\pm 2.08$ & harmonics, $3:10$ \\
1127 & GRB070508 & $2.14\pm 0.26$; $4.43\pm 0.87$ & harmonics, $1:2$ \\
1185 & GRB060906 & $4.77\pm 0.68$ & constant \\
1324 & GRB050418 & $14.70\rightarrow 4.76$ & up-chirp \\
1335 & GRB050306 & $27.97\pm 3.93$ & constant \\
\hline
\end{tabular}
\label{table1}
\end{center}
\end{table}

\section{Interpretation}

Time scales related to the period on the innermost stable circular orbit (ISCO) as the edge of the accretion disk, $T_{\rm ISCO} = 12\pi\sqrt{6}GM_{\bullet}/c^3$ \citep{hartle}, for a representative black hole (BH) mass $M_{\bullet}=10M_{\odot}$ give time scales on the order of 4.5~ms, which are shorter than the binning of the utilized LCs. If the disk is truncated at $r=kr_{\rm ISCO}$, then $T=k^{3/2}T_{\rm ISCO}$, which for $k=20\div 100$ yields $T=0.4\div 4.5$~sec. This could explain some, but not all QPOs. In turn, nonplanar relativistic orbits around a Kerr BH naturally give rise to $T=1\div 10$~sec \citep{rana20}.

Lense-Thirring precession matches the detected QPOs for a truncated disk with $k\gtrsim 30$ \citep{ingram09}, hence might be a viable possibility. On the contrary, epicyclic modes of accretion disk oscillations give rise to QPOs with $T\sim 10$~ms \citep{kotrlova20}, again shorter than the employed binning.

The tidal disruption model predicts 
that inhomogeneities with density $\rho$ in the disk will be stretched and disrupted at the Roche limit, and eventually lead to modulation with a period $T_{\rm TD}\sim \left( G\rho \right)^{-1/2}$ \citep{cadez08,kostic09,torok11}. For rocky material, this leads to $T_{TD}\sim 1650$~sec, but a compact object (e.g., a neutron star) gives $T_{TD}\sim 10$~sec; however, this is unlikely to be a universal explanation.

Up-chirps could originate from forced perturbations within a magnetized disk \citep{petri05}, but down-chirps would require a different mechanism, e.g., helical jets \citep{mohan15}. On the other hand, magnetically choked accretion flows lead to $T\sim 70GM_{\bullet}/c^3\to\,\,\sim 4$~ms \citep{mckinney12} -- even shorter than other scenarios considered and rejected above. Finally, oscillations of the shock front can give rise to QPOs with hertz and sub-hertz frequencies, so they could explain some of the QPOs in GRBs as well \citep{iyer15,palit19}.

Therefore, the possible mechanisms giving rise to QPOs in long GRBs, based on the timing constraints, are: (i) nonplanar relativistic orbits around newly formed BHs, (ii) Lense-Thirring effect, and (iii) shock oscillations.

\acknowledgements{Funding from the National Science Center through the Sonata grant No. 2021/43/D/ST9/01153 is acknowledged.}

\bibliographystyle{ptapap}
\bibliography{ptapapdoc}

\end{document}